\documentclass[prb,twocolumn,showpacs,preprintnumbers,superscriptaddress,amssymb]{revtex4}

\usepackage{graphicx}
\usepackage{dcolumn}
\usepackage{bm}
\usepackage{color}

\begin{document}

\title{Temperature dependent Eu 3$d$-4$f$
X-ray Absorption and Resonant 
Photoemission Study of  the Valence Transition in EuNi$_{2}$(Si$_{0.2}$Ge$_{0.8}$)$_{2}$}

\author{K. Yamamoto}

\affiliation{Graduate School of Engineering, Osaka Prefecture University, Sakai, Osaka 599-8531, Japan}
\affiliation{RIKEN/SPring-8, Mikazuki-cho, Sayo-gun, Hyogo 679-5148, Japan}

\author{K. Horiba}

\affiliation{RIKEN/SPring-8, Mikazuki-cho, Sayo-gun, Hyogo 679-5148, Japan}

\author{M. Taguchi}

\affiliation{RIKEN/SPring-8, Mikazuki-cho, Sayo-gun, Hyogo 679-5148, Japan}
 
\author{M. Matsunami}

\affiliation{RIKEN/SPring-8, Mikazuki-cho, Sayo-gun, Hyogo 679-5148, Japan}

\author{N. Kamakura}

\affiliation{RIKEN/SPring-8, Mikazuki-cho, Sayo-gun, Hyogo 679-5148, Japan}
 
\author{A. Chainani} 

\affiliation{RIKEN/SPring-8, Mikazuki-cho, Sayo-gun, Hyogo 679-5148, Japan}

\author{Y. Takata} 

\affiliation{RIKEN/SPring-8, Mikazuki-cho, Sayo-gun, Hyogo 679-5148, Japan}

\author{K. Mimura} 

\affiliation{Graduate School of Engineering, Osaka Prefecture University, Sakai, Osaka 599-8531, Japan}

\author{M. Shiga}

\affiliation{Department of Materials Science and Engineering, Kyoto University, Kyoto 606-8501, Japan}

\author{H. Wada}

\affiliation{Department of Materials Science and Engineering, Kyoto University, Kyoto 606-8501, Japan}

\author{Y. Senba}

\affiliation{JASRI/SPring-8, Mikazuki-cho, Sayo-gun, Hyogo 679-5198, Japan}

\author{H. Ohashi}

\affiliation{JASRI/SPring-8, Mikazuki-cho, Sayo-gun, Hyogo 679-5198, Japan}
 
\author{S. Shin}

\affiliation{RIKEN/SPring-8, Mikazuki-cho, Sayo-gun, Hyogo 679-5148, Japan}
\affiliation{The Institute for Solid State Physics, The University of Tokyo, Kashiwa, Chiba 277-8581, Japan}

\date{\today}

\begin{abstract}

We study the mixed valence transition ($T$$_{v}$ $\sim$80 K) in EuNi$_{2}$(Si$_{0.2}$Ge$_{0.8}$)$_{2}$
using Eu 3$d$-4$f$ X-ray absorption spectroscopy (XAS) and resonant photoemission 
spectroscopy (RESPES). The Eu$^{2+}$ and Eu$^{3+}$ main peaks show a giant resonance and the spectral features match 
very well with atomic multiplet calculations.
The spectra show dramatic temperature ($T$)-dependent changes
over large energies ($\sim$10 eV) in RESPES and XAS.
The observed non-integral mean valencies of $\sim$2.35 $\pm$ 0.03 ($T$ = 120 K) and $\sim$2.70 $\pm$ 0.03 ($T$ = 40 K) indicate homogeneous mixed valence above and below $T$$_{v}$.
The redistribution between Eu$^{2+}$$4f^7$+$[spd]^0$ and Eu$^{3+}$$4f^6$+$[spd]^1$ states is
attributed to a hybridization change coupled to a Kondo-like volume collapse.

\end{abstract}
\maketitle

An important issue of enduring interest in $f$-electron systems which exhibit mixed-valence (MV) transitions
 and the related Kondo effect is the role of purely ionic-like states compared to delocalized or
 hybridized states.\cite{varma,free,dallera1,rueff,haule} For example, 
in SmS, which exhibits a pressure dependent 
MV transition\cite{jaya}, the relevant ionic states are Sm$^{2+}$ $4f^6$  and Sm$^{3+}$ $4f^5$ states. It is clear that such a  transition
requires the role of the $5d$ electrons of Sm, or more generally, the [$spd$] electrons of the conduction band.\cite{varma} The necessary condition of charge balance for SmS would indicate that
the transition involves Sm$^{2+}$ $4f^6$, and the Sm$^{3+}$ $4f^5$ + $[spd]^1$ electron configurations.
The MV transition then occurs between two stable states, each defined 
by relative contributions of the $4f^{n}$ and $4f^{n-1}$+$[spd]^1$ configurations. 
Each state can be a homogeneous MV state
having the same non integral 
valence at each site, due to a quantum 
mechanical mixing of the relevant configurations.\cite{croft} In contrast, a static or inhomogeneous MV state is one in which electron configurations are different at different sites, representing one specific electron configuration at a site. While the MV in
$f$-electron systems are often homogeneous, there do exist exceptions.\cite{varma,felser}

Many $f$-electron systems exhibit a MV transition induced by temperature ($T$), magnetic field and/or pressure. These include the $\alpha$-$\gamma$ transition in Ce metal,\cite{varma} the pressure-induced transitions in SmS,\cite{jaya} and TmTe,\cite{matsu} the $T$-dependent transitions in YbInCu$_{4}$,\cite{felner} Tm-monochalcogenides,\cite{clay} as well as
Eu-based intermetallics, EuPd$_{2}$Si$_{2}$,\cite{sampath} Eu(Pd$_{1-x}$Au$_{x}$)$_{2}$Si$_{2}$,\cite{segre} and EuNi$_{2}$(Si$_{1-x}$Ge$_{x}$)$_{2}$.\cite{wort} 
Among $T$-induced transitions, the Eu systems exhibit the largest change in valency, $\Delta$$v$ $\sim$ 0.3-0.5.\cite{sampath,segre,wort} And of these,
EuNi$_{2}$(Si$_{1-x}$Ge$_{x}$)$_{2}$ has been extensively studied
to show $T$-,\cite{wort,wada1} magnetic field-,\cite{wada1} and pressure-\cite{wada2}
induced valence transitions.
By tuning composition [x in EuNi$_{2}$(Si$_{1-x}$Ge$_{x}$)$_{2}$], the transition is observed to be first-order like for compositions close to $x$ = 0.8, with a hysterisis as a  function of $T$, pressure and magnetic field.\cite{wada1,wada2,wada3} The MV transition in EuNi$_{2}$(Si$_{0.2}$Ge$_{0.8}$)$_{2}$ has thus been investigated across the critical $T$ ($T$$_{v}$) of $\sim$80 K by magnetic susceptibility, high-energy bulk-sensitive Eu $L$-edge X-ray absorption spectroscopy (XAS), and X-ray diffraction to show that the transition is accompanied by a Kondo-like volume collapse across $T$$_{v}$.\cite{wort,wada1,wada2,wada3}

XAS and resonant photoemission spectroscopy (RESPES) are important techniques for studying the electronic structure (ES) of  $f$-electron systems.\cite{gunnar,fuggle,allen} 
In XAS applied to a solid, a core electron of a particular site or element
is excited to an empty state, and hence, it probes site-specific angular momentum projected unoccupied states of a solid.\cite{degroot} RESPES is a complementary technique which probes the resonantly enhanced partial occupied density of states (DOS) of a solid.\cite{allen} These techniques provide important insights into the physical properties of strongly correlated materials, including MV, Kondo effect, heavy fermion behavior, etc. 
However, recent studies using ultraviolet photoemission spectroscopy (PES) of MV systems
revealed modifications of the surface ES compared to the bulk.
\cite{claessen,moore,ray}
While signatures of $T$-dependent MV are observed, the mean valence estimated from these measurements are incompatible with bulk thermodynamic studies.
Significantly, the important role of hard X-ray (HX: h$\nu$ $\sim$ 3 to 8 keV) PES in general,\cite{braico,kobayashi,takata,drube,
dallera2,torelli,sato2} as well as soft X-ray (SX: h$\nu$ $\sim$ 1 to 1.5 keV) PES\cite{claessen} and
RESPES\cite{sekiyama,cho} of $f$-electron systems, has been reiterated
for studying bulk character ES. 

In this work we study $T$-dependence of the MV transition in EuNi$_{2}$(Si$_{0.2}$Ge$_{0.8}$)$_{2}$ using XAS and RESPES across the Eu 3$d$-4$f$ threshold. We observe a giant resonance of  Eu$^{2+}$ and Eu$^{3+}$ main peaks and dramatic $T$-dependent changes in the XAS and RESPES data. The mean valence estimated from the data are consistent with bulk sensitive results, 
indicating non-integral homogeneous mean valencies of 2.35 $\pm$ 0.03 (above $T$$_{v}$) and 2.70 $\pm$ 0.03 (below $T$$_{v}$).
The XAS data are analyzed using atomic multiplet calculations for Eu$^{2+}$ and Eu$^{3+}$ states. The RESPES valence band spectra as a function of energy are also consistent with atomic calculations. The $T$-dependent transition redistributes occupancies of
 Eu$^{2+}$ $4f^7$+$[spd]^0$ and Eu$^{3+}$$4f^6$+$[spd]^1$ configurations, attributed to
a hybridization change coupled to a Kondo-like volume collapse.

EuNi$_{2}$(Si$_{0.2}$Ge$_{0.8}$)$_{2}$ polycrystalline samples were prepared by melting stoichiometric amounts of constituent elements in an argon furnace.\cite{wada1} EuNi$_{2}$(Si$_{0.2}$Ge$_{0.8}$)$_{2}$ was characterized to exhibit $T$$_{v}$ $\sim$80 K by magnetic susceptibility. The sample was single phase with the ThCr$_{2}$Si$_{2}$-type structure, as confirmed by X-ray diffraction. The Eu 3$d$-4$f$ XAS and RESPES experiments were performed at SX undulator beam line BL17SU of SPring-8 using a grazing incidence monochromator. The XAS measurements were carried out by recording sample drain current as a function of photon energy. The RESPES experiments used a hemispherical high energy-resolution electron analyzer, SCIENTA SES-2002. 
The total energy resolution at the 3$d$-4$f$ threshold was about 300 meV and the vacuum was 4$\times$10$^{-8}$ Pa. 
A clean surface was obtained by fracturing at 40 K. The measurements were performed at 120 K and 40 K in a $T$-cycle to confirm $T$-dependent changes.

\begin{figure}[t]
\begin{center}
\includegraphics[width=0.45\textwidth]{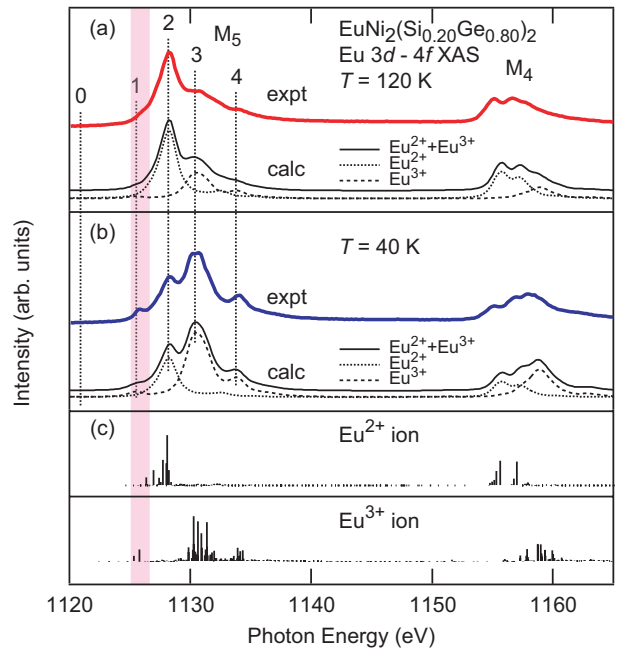}
\end{center}
\caption{(Color online) Comparison of the Eu 3$d$-4$f$ XAS experimental spectra with calculated spectra at (a) $T$ = 120 K and (b) 40 K. The spectral intensities change strongly across $T$$_{v}$ $\sim$80 K, over large energy scales. The spectra are derived from
intra-atomic multiplet excitations of Eu$^{2+}$ and Eu$^{3+}$ ions by broadening the discrete energy states (Fig. 1(c)). }
\label{fig1}
\end{figure}

Figures 1(a) and (b) show the Eu 3$d$-4$f$ XAS ($M$$_{4,5}$-edge) spectra of 
EuNi$_{2}$(Si$_{0.2}$Ge$_{0.8}$)$_{2}$ obtained at a sample $T$ of 120 K (above $T$$_{v}$) and 40 K (below $T$$_{v}$), respectively. The spectra show multiple peak structures and the intensities of the peaks show large changes over a large energy range ($\sim$10 eV in the $M$$_5$ region), as a function of $T$. 
 In order to identify the character of the various features, we calculated the Eu 3$d$-4$f$ XAS spectra and compared them with experimental results, as shown in Fig. 1. 
The calculations are atomic mutiplet calculations for the Eu$^{2+}$ and Eu$^{3+}$ free ions.\cite{cowan} 
The Slater integrals and spin-orbit coupling constants are calculated by the Hartree-Fock method with relativistic corrections. As usual, the Slater integrals are reduced to 80\%.\cite{cowan} The Eu$^{2+}$ and Eu$^{3+}$ discrete energy states are plotted as a bar diagram in Fig. 1(c). 
The discrete energy states were broadened by a Gaussian for the experimental resolution and by a Lorentzian to represent the lifetime broadening. The calculated spectra (Fig. 1(a) and (b))
show very good match with all the divalent and trivalent multiplet features, confirming that intra-atomic multiplet effects account for the observed features. The spectral intensities required to match the experiment indicate non-integral homogeneous mean valencies of 2.35 $\pm$ 0.03 at 120 K, which changes to 2.70 $\pm$ 0.03 at 40 K. 
These values match with the mean valence estimated from bulk-sensitive $L$-edge XAS\cite{wort} and HX-PES 
of Eu $3d$ core levels\cite{braico},
which showed a mean valence change from $\sim$2.75 to $\sim$2.40.\cite{yamamoto}
An important point to note is that the lowest unoccupied states with significant intensity
for Eu$^{2+}$ and Eu$^{3+}$ configurations
are close in energy and constitute the feature labelled 1 shaded region. This seems, at first glance, quite surprising because the 
the Eu$^{2+}$ and Eu$^{3+}$ features are well-separated in the core levels\cite{yamamoto}
and valence band (discussed below) due to the strong on-site Coulomb correlations in the $f$ states ( $U$$_{ff}$$\ge$5 eV). However, in the atomic multiplet approximation for XAS, if we consider the 3$d$-4$f$ excitations 
to be dominated by $U$$_{ff}$ and the core hole attraction $U$$_{fc}$, the 3$d$-4$f$ excitation
energy difference between Eu$^{2+}$ and Eu$^{3+}$ is obtained to be $U$$_{ff}$-$U$$_{fc}$. Since 
$U$$_{ff}$$\sim$1.2$U$$_{fc}$, it suggests that a lower energy scale, 
such as hybridization between $f$ and conduction band $[spd]$ states, can play a major role in the MV
transition.

\begin{figure}[ht]
\begin{center}
\includegraphics[width=0.38\textwidth]{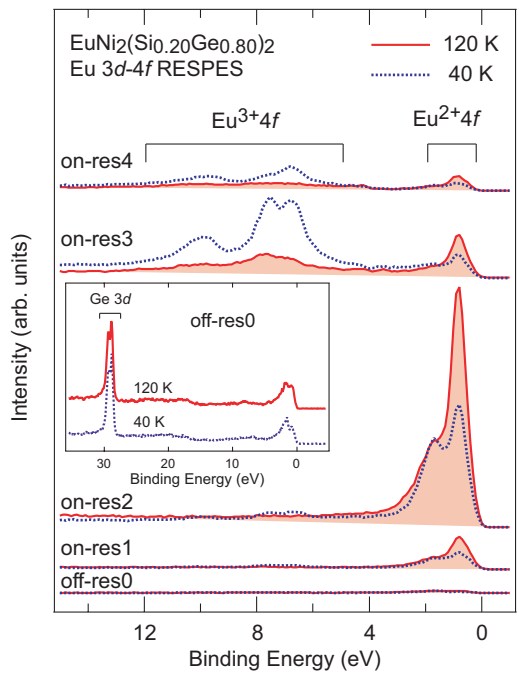}
\end{center}
\caption{(Color online) The $T$-dependent Eu 3$d$-4$f$ RESPES spectra at incident photon energies marked 0-4 in Fig.1. Inset shows normalization procedure using off-resonance (off-res0) spectra. The resonantly enhanced
features originate from partial $4f$ DOS of Eu$^{2+}$ and Eu$^{3+}$ ions. Note absence of T-dependence
in Eu$^{2+}$ surface-state at 1.7 eV BE (on-res2). }
\label{fig2}
\end{figure}

Fig. 2 shows the Eu 3$d$-4$f$ valence band RESPES obtained at $T$ = 120 and 40 K for 
energies labelled 0-4 in Fig.1 [corresponding to 
incident photon energies of  h$\nu$ = 1121.0 eV (off-resonance), 1125.7 eV, 1128.2 eV, 1130.2 eV and 1133.9 eV]. 
All the spectra are normalized to the Ge $3d$ shallow core levels which do not change shape as a function of $T$ and incident  photon energy,
as shown in inset for the off-resonant spectra over a larger energy scale (0-35 eV). In the off-reasonant spectra, the feature at 0.8 eV binding energy (BE) is the bulk Eu$^{2+}$ 4$f$ state, and the feature at 1.5 eV BE, composed of the Ni 3$d$ and the surface Eu$^{2+}$ 4$f$ states, has higher intensity.\cite{ray,kino} This results in effectively masking the $T$-dependent changes in off-resonant spectra as the Eu$^{2+}$ 4$f$ state does not participate in the MV transition.\cite{ray,martensson}
In contrast, the on-resonant spectra at energies on-res1-4 show dramatic changes in spectral features and intensity as a function
of $T$ and incident photon energy. We discuss all the spectral features and peak assignments using
atomic multiplet calculations for the RESPES data at the energies on-res1-4, shown in Fig. 3.

\begin{figure}[ht]
\begin{center}
\includegraphics[width=0.45\textwidth]{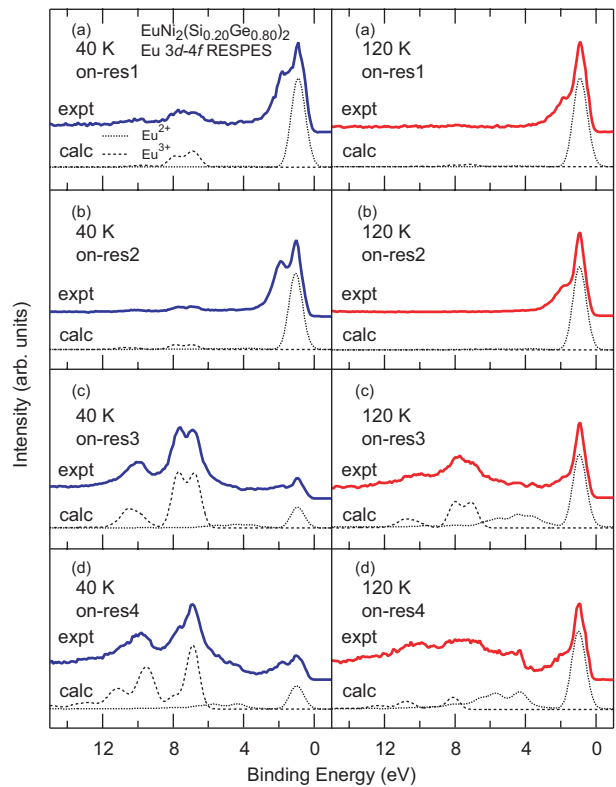}
\end{center}
\caption{(Color online) Comparison of the Eu 3$d$-4$f$ RESPES experimental spectra
with atomic multiplet calculations at incident photon energies from on-res1 to on-res4, at 40 K (left panel)
and  120 K (right panel).}
\label{fig3 (a-d)}
\end{figure}

The spectra at the photon energy on-res1 is enhanced for the bulk Eu$^{2+}$ feature at 
0.8 eV and surface feature at 1.7 eV BE (Fig. 2). The bulk feature shows a further clear enhancement at 120 K compared to 40 K, while the surface Eu$^{2+}$ feature shows negligible change in spectral intensity with $T$. Weak features between 6-8 eV BE
indicate $T$-dependent changes between 120 and 40 K (Fig. 3(a)), with strongly reduced intensity at 120 K. The atomic mutiplet calculations (dotted lines, Fig. 3(a))
confirm the Eu$^{2+}$ character of the feature at 0.8 eV and the Eu$^{3+}$ character between 6-8 eV with very weak features at higher energies.

At the energy on-res2,
the  Eu$^{2+}$ features are strongly enhanced, as this energy corresponds to the main peak of Eu$^{2+}$  character in XAS (Fig. 1). It is a giant resonant enhancement because the intensity increase at the bulk Eu$^{2+}$ peak is $\sim$140 times compared to the off-resonance data at 120 K.\cite{kunz2}
At this energy, the features between 6-8 eV BE which originate in Eu$^{3+}$ 
states, also show clear $T$-dependence: the spectral intensity nearly vanishes at 120 K but is observed at 40 K.
This observation confirms that the $T$-dependence is due to intrinsic mixed valency changes and 
not due to oxidation at high $T$, as oxidation would lead to an increase in Eu-trivalency at 120 K.  
The calculations match the bulk Eu$^{2+}$ and Eu$^{3+}$ features.

On increasing the energy to on-res3, the Eu$^{3+}$ features are strongly enhanced with a simultaneous
reduction of the Eu$^{2+}$ features. This energy being the main peak of Eu$^{3+}$ derived states in XAS, 
the weaker Eu$^{3+}$ multiplets are also significantly enhanced upto a BE of nearly 12 eV (Fig. 2 
and Fig. 3(c)). The Eu$^{3+}$ main peak also shows a giant resonance: $\sim$100 times increase compared to off-resonance at 40 K. Note also the strong $T$-dependence in the Eu$^{3+}$ features at this energy (Fig. 2).
Interestingly, while the calculations match the Eu$^{2+}$ and Eu$^{3+}$ derived features as obtained
at lower energies, a new feature is obtained between the main Eu$^{2+}$ and Eu$^{3+}$ features (Fig. 3(c)).
This is seen as a weak tail between 3-6 eV i.e. at BEs lower than the main
 Eu$^{3+}$ peak. This is assigned to the intermediate
spin-flip satellites, as is known from earlier work.\cite{kunz} 
The spin-flip states (4$\it{f}$$^{5\uparrow,1\downarrow}$)  
lie between the  
(4$\it{f}$$^{6\uparrow}$) and 
(4$\it{f}$$^{5\uparrow}$) photoemission 
final states of Eu$^{2+}$  and Eu$^{3+}$ initial states, respectively\cite{kunz} 
and also exhibit $T$-dependence.\cite{kino,ray} 
At the energy on-res4, an overall reduction of the spectral intensities
of the Eu$^{2+}$ and Eu$^{3+}$ is observed, and the features are in agreement with the calculations (Fig. 3(d)).
The Eu$^{2+}$ surface state which was nearly absent at on-res3 energy is again recovered as a weak feature.

The success of the present XAS results in revealing the MV transition consistent with bulk sensitive $L$-edge XAS and HX-PES, is because we fractured the surface instead of scraping it.\cite{kino,yamamoto} Hence, the intrinsic ES of a strongly correlated $f$ electron system undergoing a MV transition as a function of $T$ can be measured across the 3$d$-4$f$ threshold. The energy dependent 
RESPES reveal the partial $4f$ DOS in the valence band. It is surprising that the 3$d$-4$f$
XAS and RESPES data can all be explained with atomic multiplet calculations, above and below $T$$_{v}$,
because it can't explain the MV transition.
Since the MV transition is coupled to
a Kondo-like volume collapse in the low-$T$ phase, as is known from diffraction studies,\cite{wort,wada3} the volume collapse must increase hybridization between the ionic $f$ states and the conduction band $[spd]$ states. 
The change in occupancies is attributed to changes in hybridization above and below $T$$_{v}$.

In conclusion,  the Eu 3$d$-4$f$ XAS and RESPES spectra of EuNi$_{2}$(Si$_{0.2}$Ge$_{0.8}$)$_{2}$
 at 120 K and 40 K, across the MV transition at $T$$_{v}$ $\sim$80 K are consistent with bulk sensitive measurements. 
The experimental spectra correspond nicely with calculated spectra for the Eu$^{2+}$ and Eu$^{3+}$ free ion configurations in an atomic model. The mean valence was estimated to be $\sim$2.35 $\pm$ 0.03 at 120 K and $\sim$2.70 $\pm$ 0.03 at 40 K. The redistribution between Eu$^{2+}$$4f^7$+$[spd]^0$ and Eu$^{3+}$$4f^6$+$[spd]^1$ states is
attributed to a hybridization change coupled to a Kondo-like volume collapse. Eu 3$d$-4$f$ XAS is very useful for studying the bulk intrinsic ES of strongly correlated rare earth compounds undergoing a MV transition. 

 Acknowledgements
The authors thank Prof. T. Kinoshita, Dr. Y. Taguchi, Prof. K. Ichikawa and Prof. O. Aita for very valuable discussions.

\label{}


\begin{thebibliography}{00}

\bibitem{varma} For early developments see : C. M. Varma, Rev. Mod. Phys. $\bf 48$, 219 (1976);
Valence Fluctuations in Solids, ed. L. M. Falicov, W. Hanke, and M. B. Maple (North-Holland, Amsterdam,1981);
Theoretical and Experimental Aspects of Valence Fluctuations and Heavy Fermions, ed. L. C. Gupta and S. K. Malik
(Plenum, New York, 1987).


\bibitem{free} J. K. Freericks and V. Zlatic, Rev. Mod. Phys. $\bf 75$, 1333 (2003).

\bibitem{dallera1} C. Dallera $et$ $al$., Phys. Rev. Lett. $\bf 88$, 196403 (2002); C. Dallera $et$ $al$., Phys. Rev. B $\bf 70$, 085112 (2004).

\bibitem{rueff} J.-P. Rueff $et$ $al$., Phys. Rev. Lett. $\bf 93$, 067402 (2004).

\bibitem{haule} K. Haule, V. Oudovenko, S. Y. Savrasov, and G. Kotliar, Phys. Rev. Lett. $\bf 94$, 036401 (2005).

\bibitem{jaya} A. Jayaraman, V. Narayanamurti, E. Bucher, and R. G. Maines, Phys. Rev. Lett. $\bf 25$, 368 (1970).

\bibitem{croft} M. Croft, J. A. Hodges, E. Kemly, A. Krishnan, V. Murgai, and L. C. Gupta, Phys. Rev. Lett. $\bf 48$, 826 (1982) ; J. Rohler, D. Wohlleben, G. Kaindl, and H. Balster, Phys. Rev. Lett. $\bf 49$, 65 (1982).

\bibitem{felser} C. Felser $et$ $al$., Europhys. Lett. $\bf 40$, 85 (1997).

\bibitem{matsu} T. Matsumura $et$ $al$., Phys. Rev. Lett. $\bf 78$, 1138 (1997) . 

\bibitem{felner} I. Felner and I. Nowik, Phys. Rev. B $\bf 33$, 617 (1986).

\bibitem{clay} B. P. Clayman, R. W. Ward, and J. P. Tidman, 
Phys. Rev. B $\bf 16$, 3734 (1977).

\bibitem{sampath} E. V. Sampathkumaran, L. C. Gupta,  R. Vijayaraghavan, K. V. Gopalakrishnan, R. G. Pillay, and H. G. Devare, J. Phys. C $\bf 14$, L237 (1981).

\bibitem{segre} C. U. Segre, M. Croft, J. A. Hodges, V. Murgai, L. C. Gupta, and R. D. Parks, Phys. Rev. Lett. $\bf 49$, 1947 (1982).

\bibitem{wort} G. Wortmann, I. Nowik, B. Perscheid, G. Kaindl, and I. Felner, Phys. Rev. B $\bf 43$, 5261 (1991).

\bibitem{wada1} H. Wada $et$ $al$., J. Phys.: Condens. Matter $\bf 9$, 7913 (1997).

\bibitem{wada2} H. Wada, M. F. Hundley, R. Movshovich, and J. D. Thompson, Phys. Rev. B $\bf 59$, 1141 (1999).

\bibitem{wada3} H. Wada $et$ $al$., J. Phys. Soc. Jpn. $\bf 68$, 950 (1999).

\bibitem{gunnar} O. Gunnarsson and K. Schonhammer,
Phys. Rev. B $\bf 28$, 4315 (1983).

\bibitem{fuggle} J. C. Fuggle, F. U. Hillebrecht, J.-M. Esteva, R. C. Karnatak, O. Gunnarsson, and K. Schonhammer
Phys. Rev. B $\bf 27$, 4637-4643 (1983).

\bibitem{allen} J. W. Allen, in Synchrotron Radiation Research : Advances in Surface and Interface Science, ed. 
R.Z. Bachrach, pg. 253 (Plenum, New York, 1992).
 
\bibitem{degroot} F. de Groot, Chemical Reviews $\bf 101$, 1779 (2001).

\bibitem{claessen} F. Reinert $et$ $al$., Phys. Rev. B $\bf 58$, 12808 (1998) ; S. Schmidt, S. Hufner, F. Reinert, and W. Assmus
Phys. Rev. B $\bf 71$, 195110 (2005).

\bibitem{moore} D. P. Moore $et$ $al$.,
Phys. Rev. B $\bf 62$, 16492 (2000).

\bibitem{ray} P. A. Rayjada $et$ $al$.,
Phys. Rev. B $\bf 70$, 235105 (2004).

\bibitem{braico} L. Braicovich, N. B. Brookes, C. Dallera, M. Salvietti, and G. L. Olcese, 
Phys. Rev. B $\bf 56$, 15047 (1997).

\bibitem{kobayashi} K. Kobayashi $et$ $al$., Appl. Phys. Lett. $\bf 83$, 1005 (2003).

\bibitem{takata}Y. Takata $et$ $al$., Appl. Phys. Lett. $\bf 84$, 4310 (2004).

\bibitem{drube} W. Drube, T. Eickhoff, H. Schulte-Schrepping, and J. Heuer, AIP Conference Proceedings $\bf 705$, 1130 (2004).

\bibitem{dallera2} C. Dallera $et$ $al$., Appl. Phys. Lett.
$\bf 85$, 4532 (2004).

\bibitem{torelli}
P. Torelli $et$ $al$., Rev. Sci. Instrum. $\bf 76$, 023909 (2005).

\bibitem{sato2} H. Sato $et$ $al$.,
Phys. Rev. Lett. $\bf 93$, 246404 (2004).

\bibitem{sekiyama} A. Sekiyama, T. Iwasaki, K. Matsuda, Y. Saitoh, Y. Onuki, and S. Suga, Nature $\bf 403$, 396 (2000).

\bibitem{cho} E.-J. Cho $et$ $al$.,
Phys. Rev. B $\bf 67$, 155107 (2003).

\bibitem{cowan} R. Cowan, The Theory of Atomic Structure and Spectra (University of California Press, Berkeley, 1981).

\bibitem{yamamoto} K. Yamamoto $et$ $al$., J. Phys. Soc. Jpn. $\bf 73$, 2616 (2004).

\bibitem{kino} T. Kinoshita $et$ $al$., J. Phys. Soc. Jpn. $\bf 71$, 148 (2002).

\bibitem{martensson} N. Martensson, B. Reihl, W.-D. Schneider, V. Murgai, L. C. Gupta, and R. D. Parks, Phys. Rev. B, $\bf 25$, R1446 (1982).

\bibitem{kunz2} W. Lenth, F. Lutz, J. Barth, G. Kalkoffen, and C. Kunz, Phys. Rev. Lett. $\bf 41$, 1185 (1978).

\bibitem{kunz}
F. Gerken, J. Barth, and C. Kunz, Phys. Rev. Lett. $\bf 47$, 993 (1981).




\end{thebibliography}
\end{document}